\documentclass[twocolumn,caps,showpacs]{revtex4}

\usepackage{graphicx,color}
\usepackage{latexsym}
\usepackage{amsmath,amssymb}        
\usepackage[draft=false]{hyperref}
\usepackage{mathrsfs}
\DeclareSymbolFontAlphabet{\mathrsfs}{rsfs}
\DeclareMathAlphabet{\mathcal}{OMS}{cmsy}{m}{n}

\begin{document}


\title{Accretion of a Phantom Scalar Field by a Black Hole: restrictions on the field distribution}


\author{J. A. Gonz\'alez and F. S. Guzm\'an}
\affiliation{Laboratorio de Inteligencia Artificial y Superc\'omputo. 
		Instituto de F\'{\i}sica y Matem\'{a}ticas, Universidad
              Michoacana de San Nicol\'as de Hidalgo. Morelia, Michoac\'{a}n,
              M\'{e}xico.}


\date{\today}


\begin{abstract}
Using numerical simulations we study the accretion of a phantom scalar field into a black hole and track the black hole horizon shrinking process. We integrate the amount of field left outside the black hole during the process in terms of the properties of the wave packet sent toward the black hole. Our results indicate that the accretion of thick wave packets with small wave number is inefficient and a portion of the scalar field remains outside the black hole.
\end{abstract}




\maketitle

\section{Introduction}
\label{sec:introduction}

The accretion process in black holes (BHs) is one of the main stream problem in relativistic astrophysics, involving different configurations related to disks, winds, accretion from a binary partner, each of them with different degrees of sophistication of the matter fields considered. For instance, it could be a perfect fluid coupled to General Relativity, or a plasma obeying the Magnetohydrodynamics equations, both with ideal or resistive MHD, or it could be a fluid coupled to radiation. All these scenarios have a direct connection to observations of high energy astrophysics. Various processes and detailed descriptions of them can be found in \cite{Luciano} and references therein. On the other hand there are other accretion processes that help exploring different properties of black holes or exotic matter fields around, whose existence and properties have some degree of speculation, but that appear in Cosmology as potential models of dark matter or dark energy. This set of scenarios involve regular scalar fields, phantom and ghost scalar fields, Galileon fields, etc. The motivation to study their behavior around black holes ranges from investigating whether black holes have hair or scalar wigs \cite{ccs1}, to studying the consequences of either how black holes affect these matter fields or the influence these fields have on the behavior or abundance of black holes. Within this field, a review explaining various scenarios of accretion onto black holes can be found in \cite{BabichevReview}.

Within the most intriguing fields are those that make a black hole shrink, which include scalar fields or fluid violating the weak energy condition \cite{BabichevReview}. The decrease of a black hole horizon due to the accretion of classical fluids has been studied in the past and has important implications, for example, that a BH should gradually vanish as the evolution of the universe approaches a Big Rip state \cite{Ref86}. Particular scenarios involving the accretion of phantom energy have shown that the black hole area decreases with the accretion \cite{Ref86,Ref87,Ref98}. Calculations of accretion of a scalar field violating the energy conditions indicating a BH area decrease can be found in \cite{Ref98,Ref102}, in the test field approximation. And also the decrease of the BH area through the accretion of a phantom scalar field has been confirmed in full non-linear General Relativity \cite{GG1}. The black hole area reduction by the accretion of a potentially existing field is an interesting subject, because it would be an alternative process for black hole evaporation.

On the other hand, there is a property of scalar fields fulfilling the energy conditions, that indicates the accretion of a scalar field is partial. The amount of accreted scalar field depends on certain conditions of the incident wave packet, in particular the wave number and the width of the packet. This has been studied both, in the test field approximation \cite{Urena2011} and in full General Relativity \cite{GG1}. Further studies in the test field limit indicate that a scalar field can be sustained by a black hole without being accreted \cite{ccs}.

In this paper we combine both scenarios, namely the accretion of a phantom field onto a black hole and study the amount of field accreted in terms of the spatial properties of the scalar field wave packet. We study this problem by solving the full Einstein-Klein-Gordon system of equations numerically and track the evolution of the black hole and scalar field in order to estimate the amount of accreted scalar field. 
The wave packets sent toward the black hole are spherical waves modulated with a Gaussian profile. We measure the amount of field accreted for various parameters of a wave packet, specifically the wave number and the width of the packet.

The paper is organized as follows, in Section \ref{sec:system} we present the system of equations describing the accretion process. In section \ref{sec:results} we present results for an explored region of the parameter space in terms of wave number and scalar field width packet. Finally in Section  \ref{sec:conclusions} we draw some conclusions.

\section{Description of the system}
\label{sec:system}

We solve Einstein's equations sourced by a scalar field with stress-energy tensor $T_{\mu\nu} = \partial_{\mu}\phi \partial_{\nu}\phi +\frac{1}{2}g_{\mu\nu} \partial^{\alpha}\phi \partial_{\alpha}\phi$, where $\phi$ is the phantom scalar field and where we have set the scalar field potential $V$ to zero. This field violates the null energy condition $T_{\mu\nu}k^{\mu}k^{\nu} \le 0$, where $k^{\mu}$ is a null vector. This conditions implies the violation of the weak energy condition, which in turn implies that observers along time-like trajectories could observe negative values of the energy density.

\subsection{Numerical relativity set up}

We assume the space-time to be spherically symmetric and we describe it in spherical coordinates using the appropriate 3+1 decomposed line-element \cite{GG1}

\begin{eqnarray}
ds^2 &=& -\left( \alpha^2 - \beta^r \beta^r \frac{g_{rr}}{\chi} \right)dt^2 
	 + 2\beta^r \frac{g_{rr}}{\chi}dtdr  \nonumber\\
	&+& \frac{1}{\chi} \left[g_{rr} dr^2 + g_{\theta\theta}
	(d\theta^2 + \sin^2 \theta d\varphi^2)\right], \label{eq:metric}
\end{eqnarray}

\noindent where $\chi$ is as a conformal factor relating this metric to a spacelike flat metric, $\beta^r$ is the only nonzero component of the shift vector and $\alpha$ is the lapse function.

We solve Einstein's equations as an initial value problem based on a 3+1 decomposition of the space-time. We construct the initial data for the above scalar field considering a puncture type of black hole initial data \cite{Bruegmann}. The evolution uses the GBSSN equations for spherical coordinates \cite{Brown}, which reduce to evolution equations for the conformal factor $\chi$, the conformal metric components $g_{rr}$ and $g_{\theta\theta}$, the non-zero component of the extrinsic curvature $A_{rr}$, the trace of the extrinsic curvature $K$ and the non-zero contracted conformal Christoffel symbol $\Gamma^r$, whose equations can be found in \cite{Brown,BShapiro}.

The gauge used for the evolution is the 1+log for the lapse and the $\Gamma-$driver for the non vanishing shift component explicitly found in  \cite{Alcubierre,LibroBaumgarte}. This gauge avoids the slice stretching expected during the evolution near the black hole horizon. In order to have control of the variables near the puncture we implemented a type of excision without excision method, which consists in multiplying the sources of the evolution equations by a factor $(r/(1+r))^4$ within an excised domain covering from the coordinate origin until one quarter of the black hole's apparent horizon radius.

The evolution of the scalar field obeys the Klein-Gordon equation

\begin{equation}
\Box \phi = \frac{1}{\sqrt{-g}}\partial_{\mu}
	[\sqrt{-g}g^{\mu\nu} \partial_{\nu}\phi] = -\partial_{\phi}V,
\label{eq:kg}
\end{equation}

\noindent where $g$ is the determinant of the space-time metric. As done in  \cite{GG1}, we solve 
this equation as a set of two equations for two first-order variables, $\pi=\partial_t \phi$ and $\xi = \partial_r\phi$, coupled to the evolution of the geometry of the space-time described above.

The whole set of evolution equations is solved using the method of lines with a 4th order Runge-Kutta integrator on a uniformly discretized spatial domain, considering second order accurate spatial stencils. We use the traditional upwind method in order to avoid causal disconnection in regions with non-zero shift.

\subsection{Initial data}

The initial data assume a given initial profile of the scalar field that we choose to be a time-symmetric pulse with the form of a spherical wave modulated by a Gaussian $\phi(r,0)=A\frac{\cos(kr)}{r}e^{-(r-r_0)^2/\sigma^2}$.

In order for this scalar field to be consistent with Einstein's equations it is necessary to solve the constraints at initial time. For this we assume time-symmetry initially, which immediately makes the momentum constraint to become an identity. The Hamiltonian constraint on the other hand is non-trivial and is solved numerically. For this we assume modified isotropic coordinates and solve it using a Runge-Kutta integrator as described in  \cite{GG1}. The gauge at the initial slice uses a precollapsed lapse $\alpha=(1+M/2r)^{-2}$, zero shift $\beta^r=0$ and zero time derivative of the shift.

\subsection{Space-time and black hole masses}

The first mass we are interested in monitoring is the black hole mass. For this we determine the apparent horizon mass during the evolution. We first locate the apparent horizon using the definition of a marginal trapped surface (MTS) that obeys the condition $\Theta = \nabla_i n^i + K_{ij} n^i n^j - K =0$, where $n^i$ is an outward pointing unit vector normal to the apparent horizon, $K_{ij}$ are the components of the extrinsic curvature of the spacelike hypersurface on which one calculates the MTSs, and $K$ its trace. The apparent horizon is the outermost among the MTSs.

In order to track the evolution of the apparent horizon we calculate $\Theta$ at every time step and locate the outermost zero of it at the coordinate radius $r_{AH}$ and calculate the mass of this horizon $M_{AH} = R_{AH}/2$, where $R_{AH}=\sqrt{g_{\theta\theta}/\chi}$ is the areal radius evaluated at $r_{AH}$.

\begin{figure}[htp]
\includegraphics[width=4.25cm]{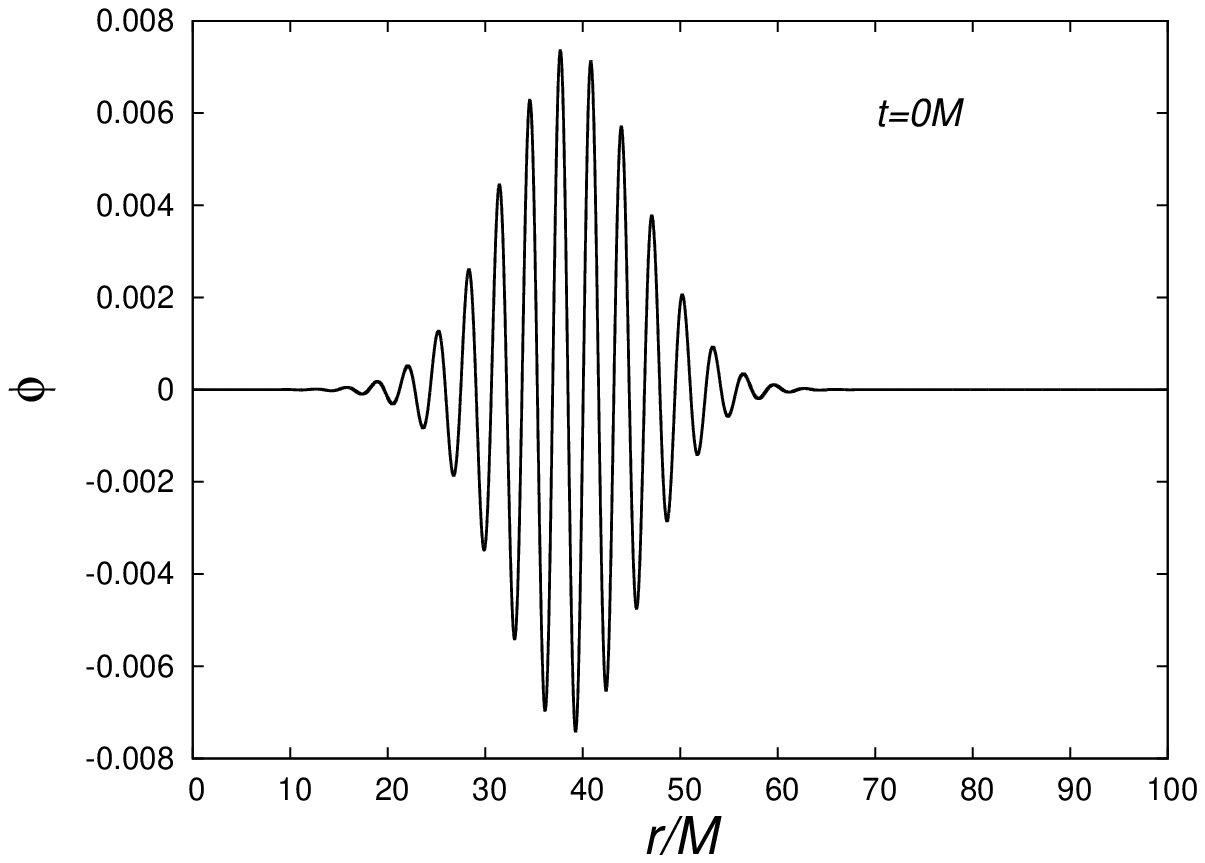}
\includegraphics[width=4.25cm]{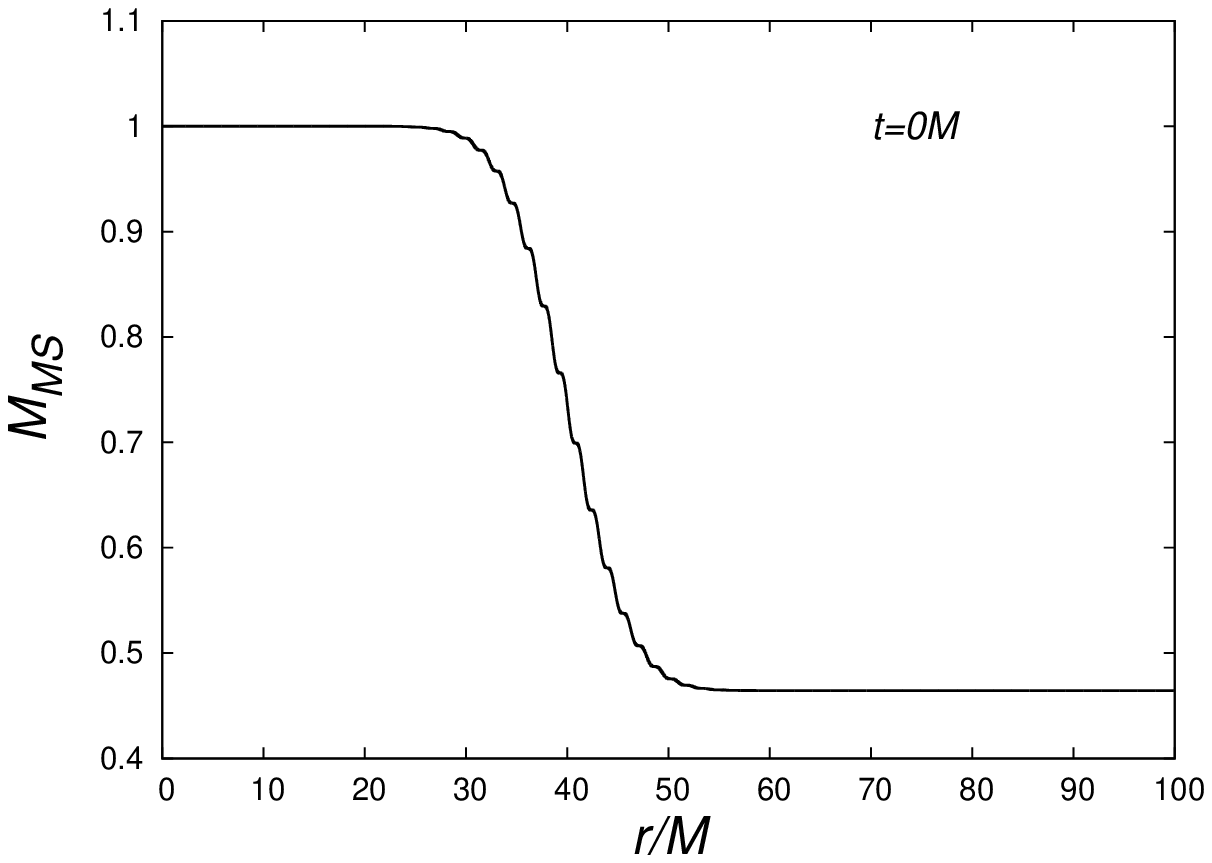}
\includegraphics[width=4.25cm]{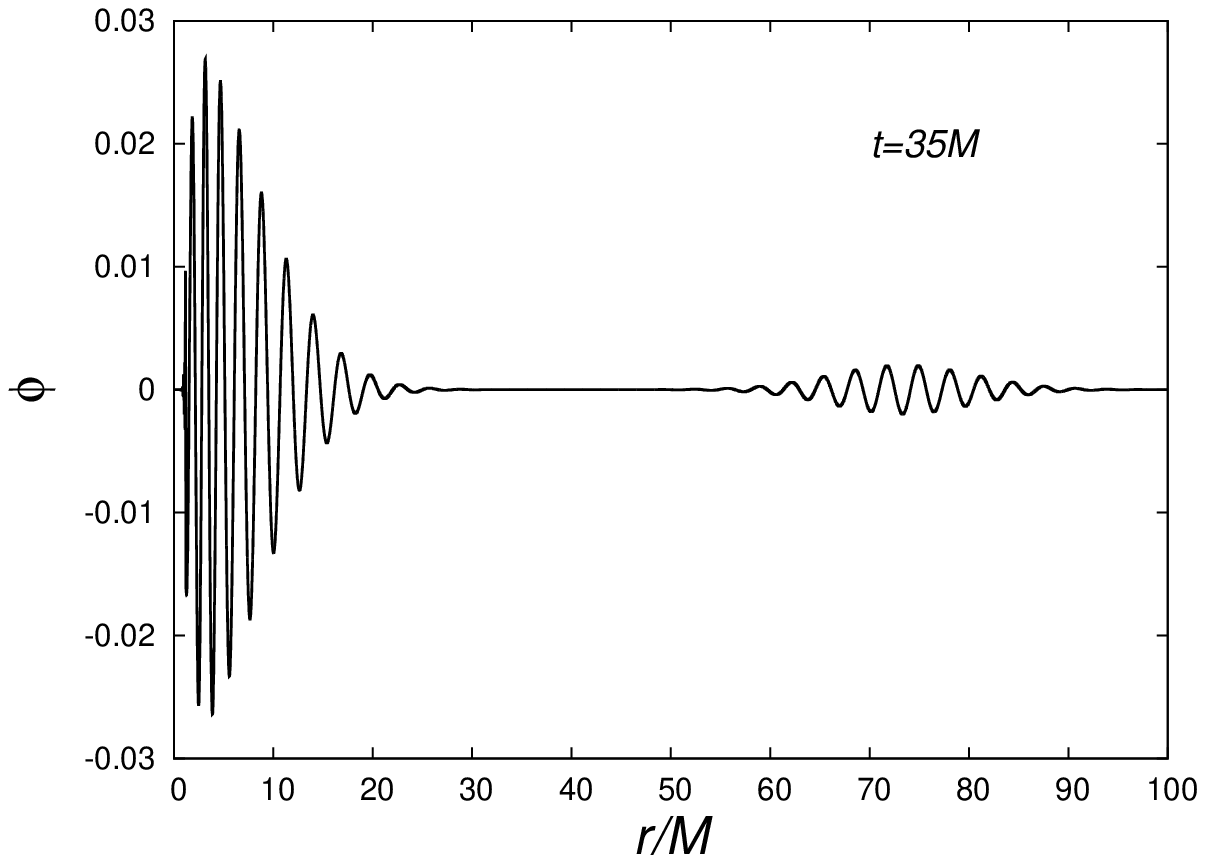}
\includegraphics[width=4.25cm]{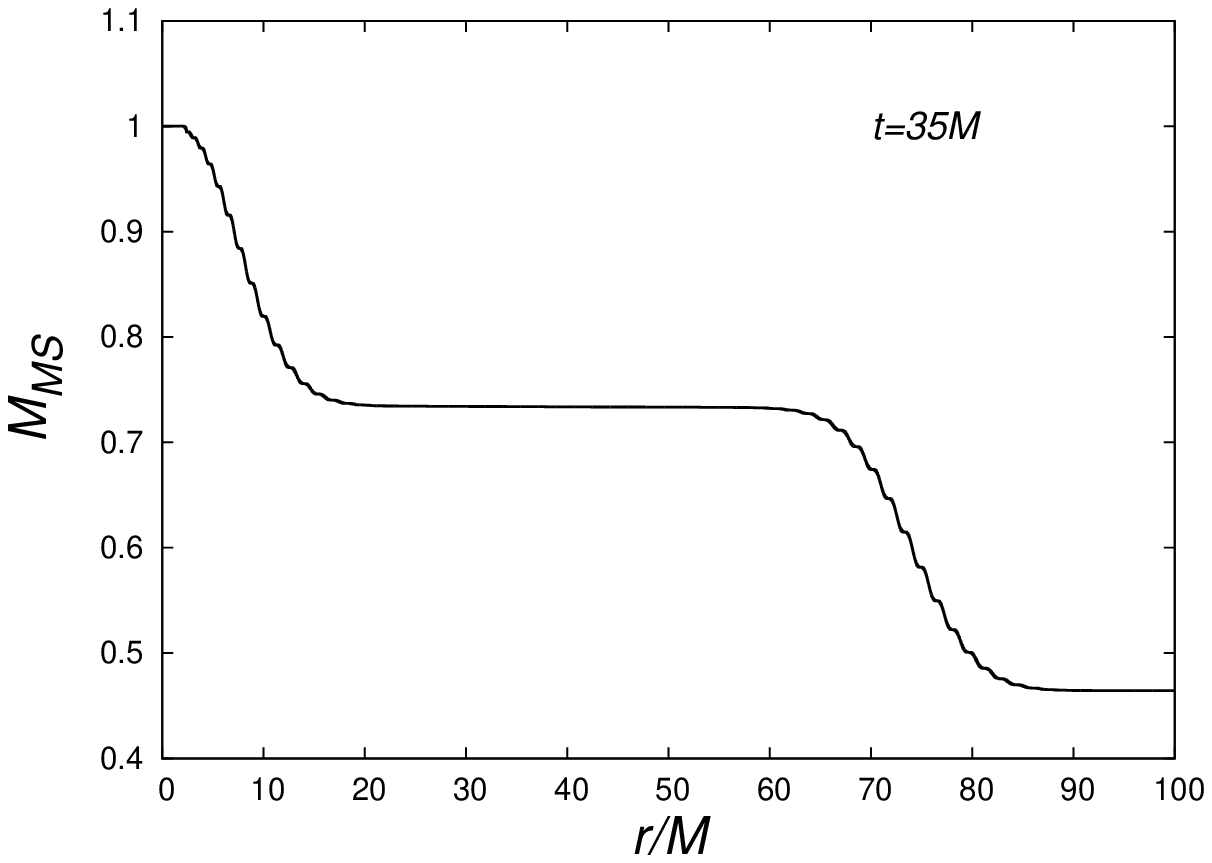}
\caption{\label{fig:massmeasure} On the left we show the scalar field and on the right the mass $M_{MS}$ at $t=0$ (top) and $t=35M$ (bottom), for the case $k=2$, $\sigma=10M$, with the pulse centered at $r_0=40M$. The mass is in units of $M$, which is the initial apparent horizon mass.}
\end{figure}

The other important mass in this analysis is the space-time mass. In the right-hand panels of Fig. \ref{fig:massmeasure} we show the Misner-Sharp mass ($M_{MS}$) of the space time as a function of the distance at two different times. The scalar field pulse is located initially at $r=40M$, thus $M_{MS}=M$ near the black hole because the only contribution to the space-time mass is the black hole. Near  $r_0=40M$ the negative contribution of the phantom scalar field can be seen and therefore $M_{MS}$ decreases until an asymptotic value. What happens once the evolution starts is that the scalar field pulse splits into two pulses. By $t=35M$ we show in the left-hand panel of Fig. \ref{fig:massmeasure} the pulses, one  moving toward the black hole and the other one moving outwards. 

The interesting pulse to investigate is the one moving toward the black hole, because it is the one that will be accreted. The ADM mass is defined for our coordinates as the value of $M_{MS}$ in the limit $r\rightarrow \infty$. This is a traditionally appropriate quantity to estimate the space-time mass. However in the present case where we want to estimate the amount of matter accreted only of the ingoing pulse, we require a measure of the space-time mass sensitive to the fact that one of the pulses has gone to infinity. For this we define our ADM mass as the Misner-Sharp mass measured at a point in the space that does not count anymore the outgoing pulse. Notice in Fig. \ref{fig:massmeasure} that by $t\sim35M$, the value of $M_{MS}$ stabilizes after the outgoing pulse has passed by, for instance at $r=40M$. Since we want to only count the mass of the ingoing pulse, this is an appropriate space-time mass measurement that we will call $M_{ADM}$. 

The value of this mass for the parameters of Fig. \ref{fig:massmeasure} is $M_{ADM}=0.73M$, where $M$ is the black hole horizon mass. In this way, we only need to measure the apparent horizon mass $M_{AH}$ during the accretion of the ingoing pulse and see whether or not it approaches $M_{ADM}$.

\section{Results}
\label{sec:results}

We explore the parameter space in wave number $k$ and width of the wave packet $\sigma$, whereas all the initial scalar field profiles we experiment with are centered at $r_0=40M$. In order to compare among the various combinations of the parameters, an essential condition of all these simulations is that $M_{ADM}=0.73$ is the same in all the cases. It can be any other value, including negative values of the space-time mass (see e.g. \cite{GG1}), but this is sufficiently arbitrary as to illustrate the development of the accretion process. In this way, for given values of $k$ and $\sigma$ the parameter we tune in order to achieve this value of $M_{ADM}$ is the amplitude $A$ of the pulse.


\begin{figure*}[htp]
\includegraphics[width=4.3cm]{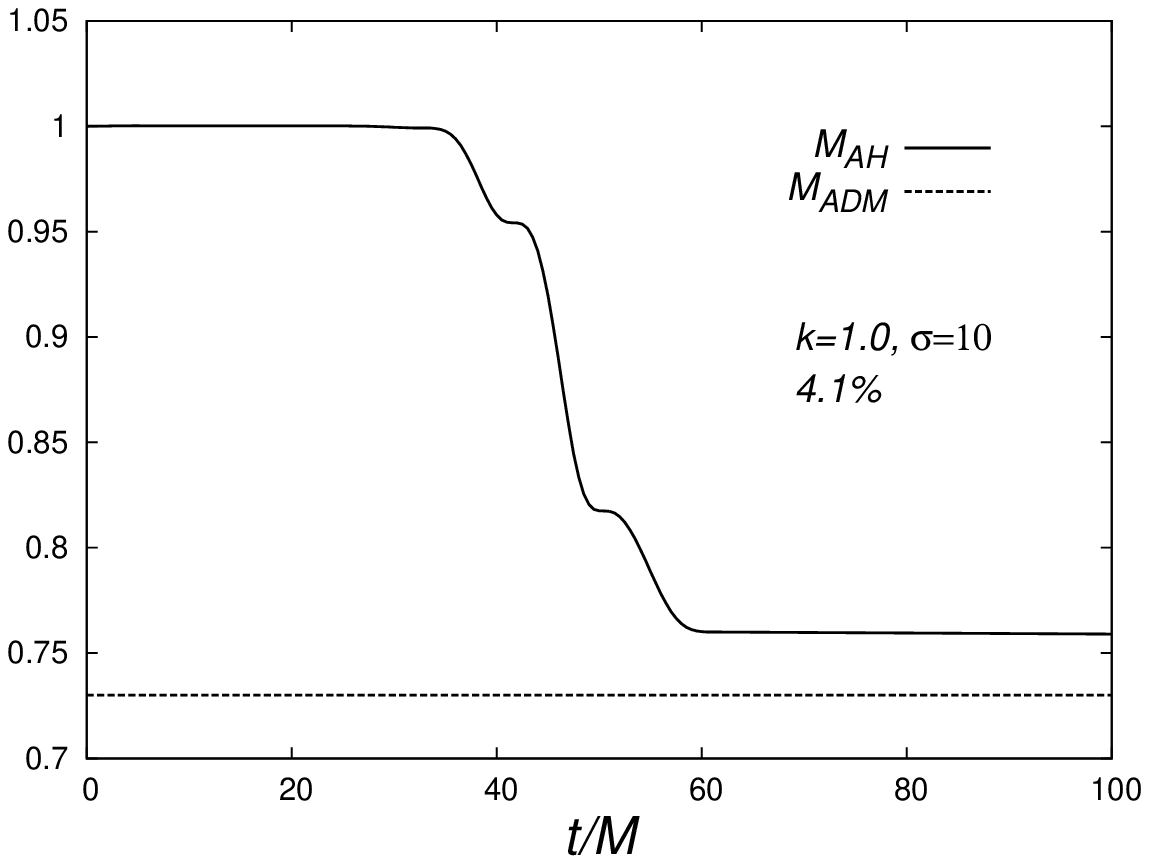}
\includegraphics[width=4.3cm]{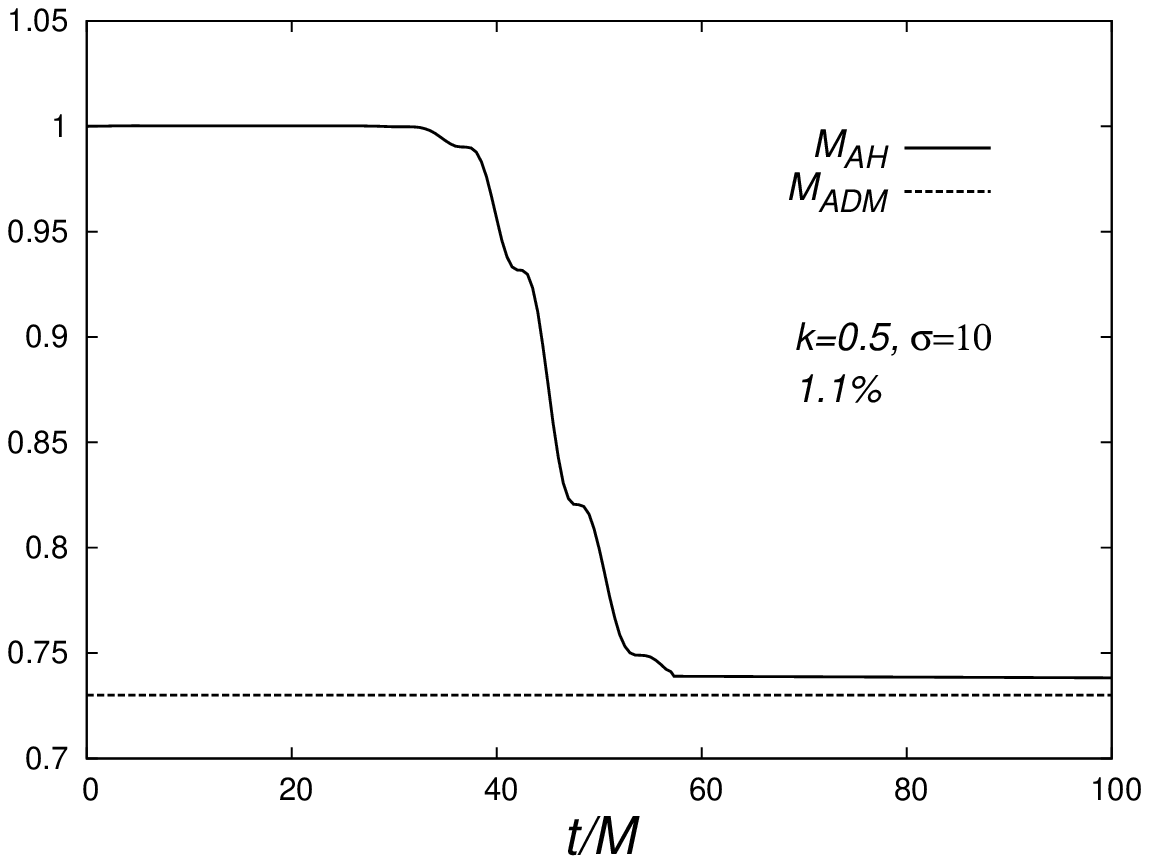}
\includegraphics[width=4.3cm]{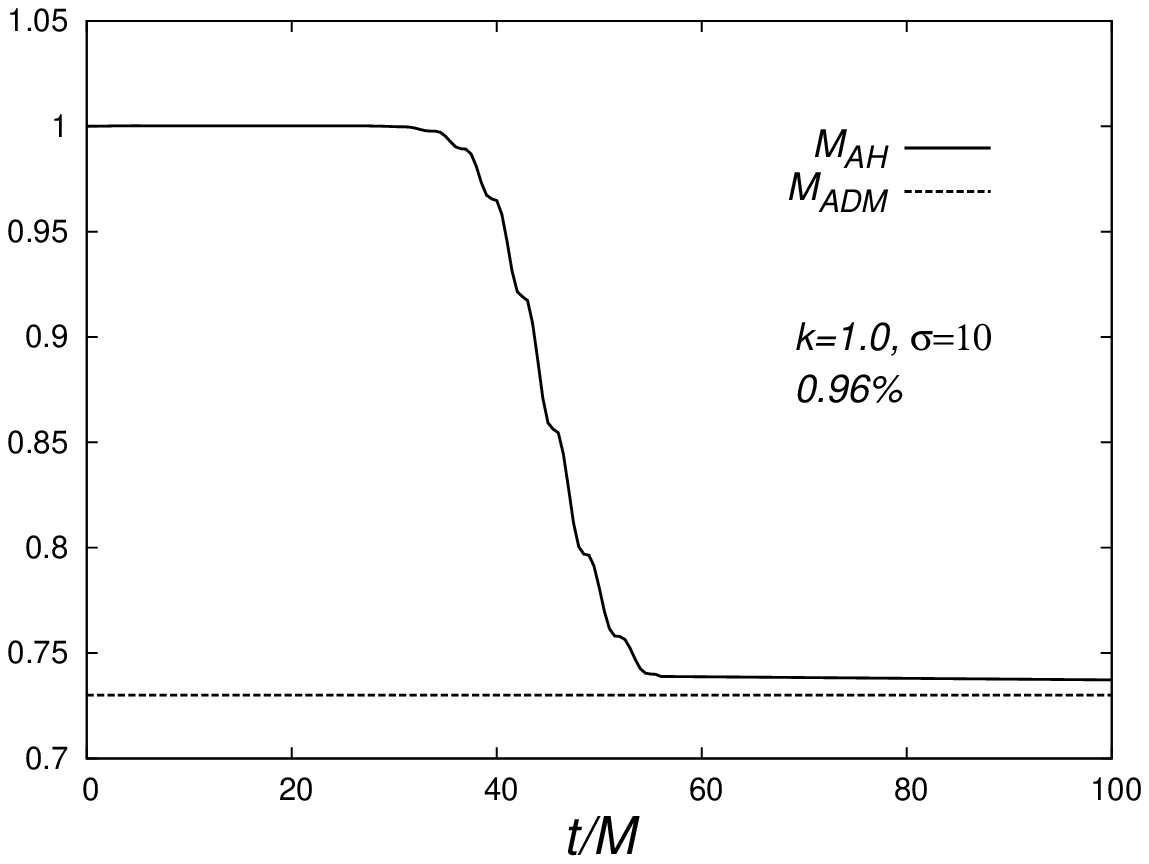}
\includegraphics[width=4.3cm]{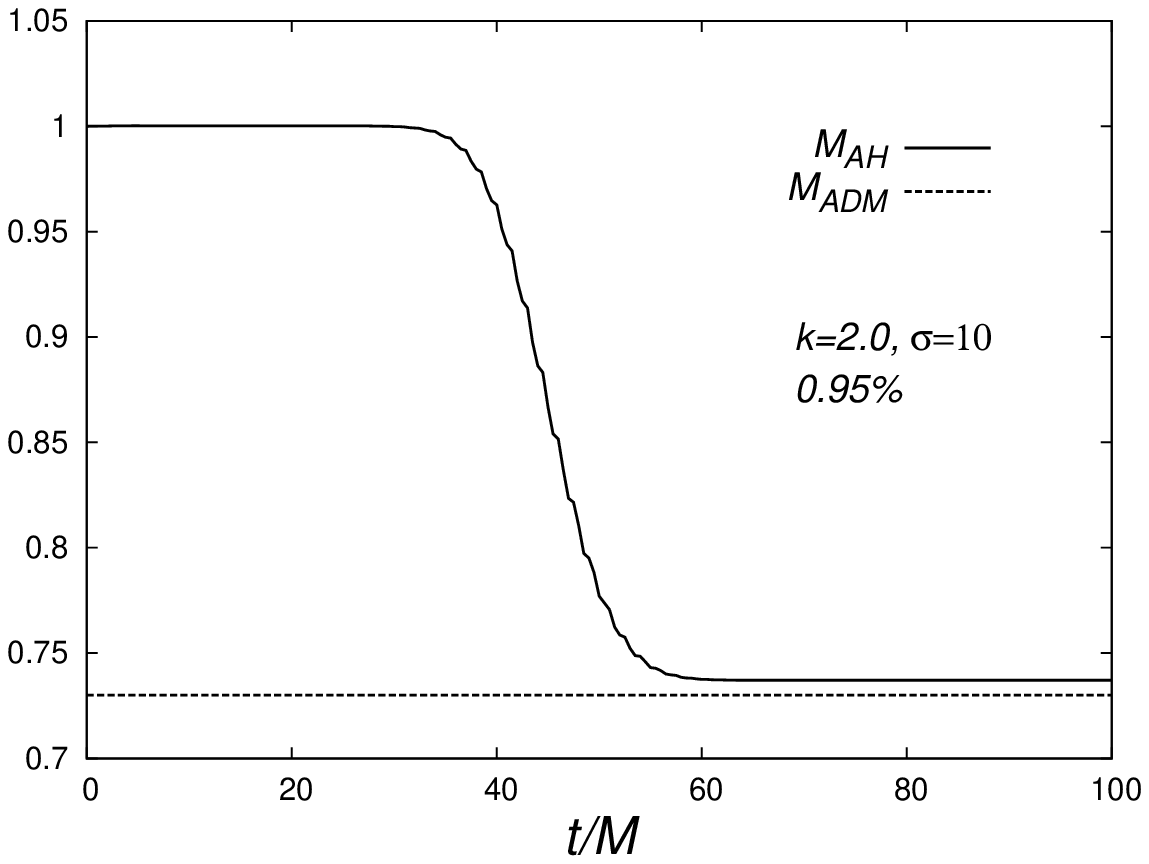}
\includegraphics[width=4.3cm]{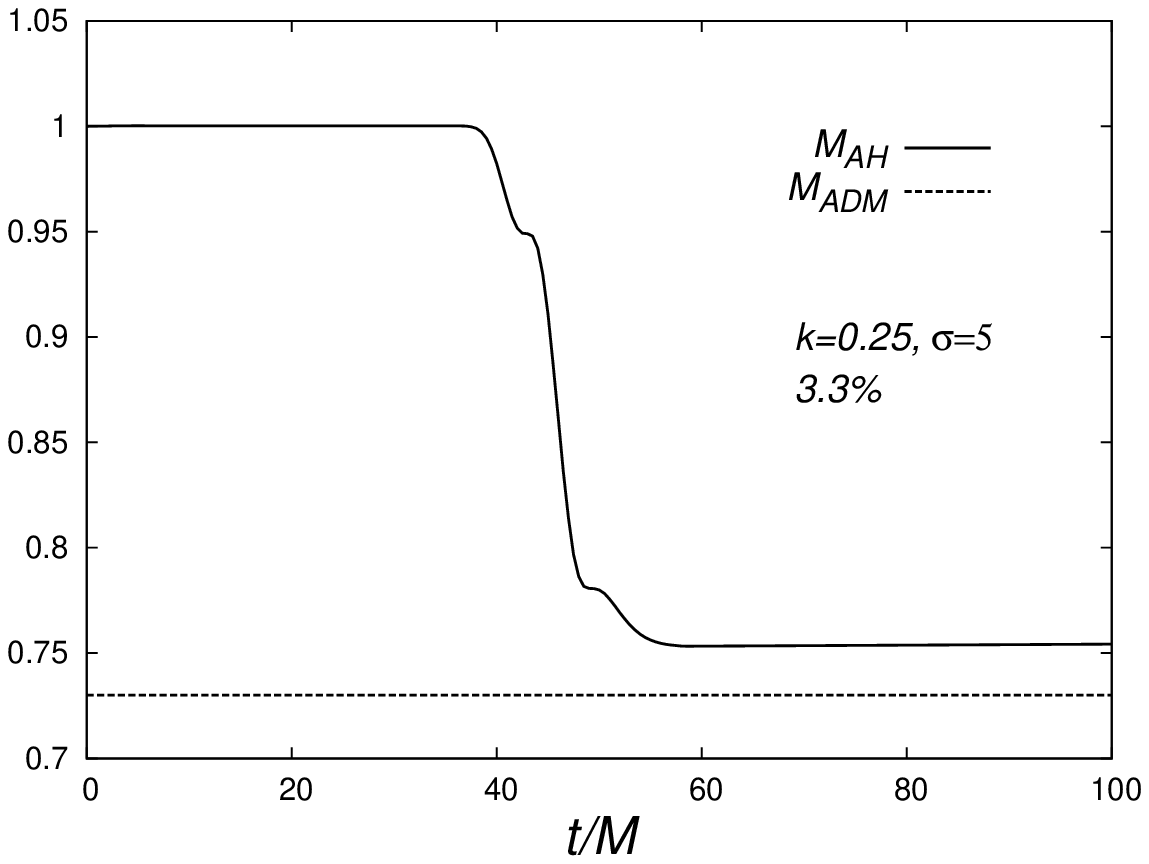}
\includegraphics[width=4.3cm]{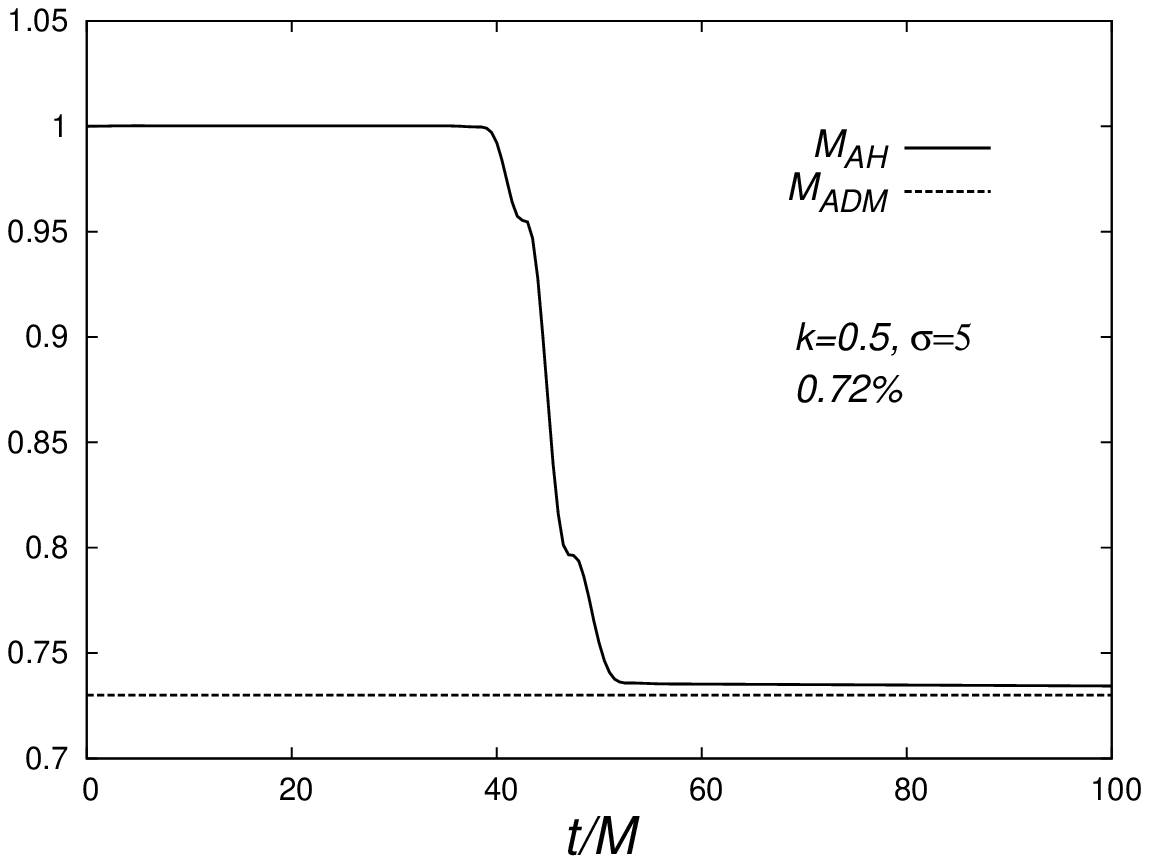}
\includegraphics[width=4.3cm]{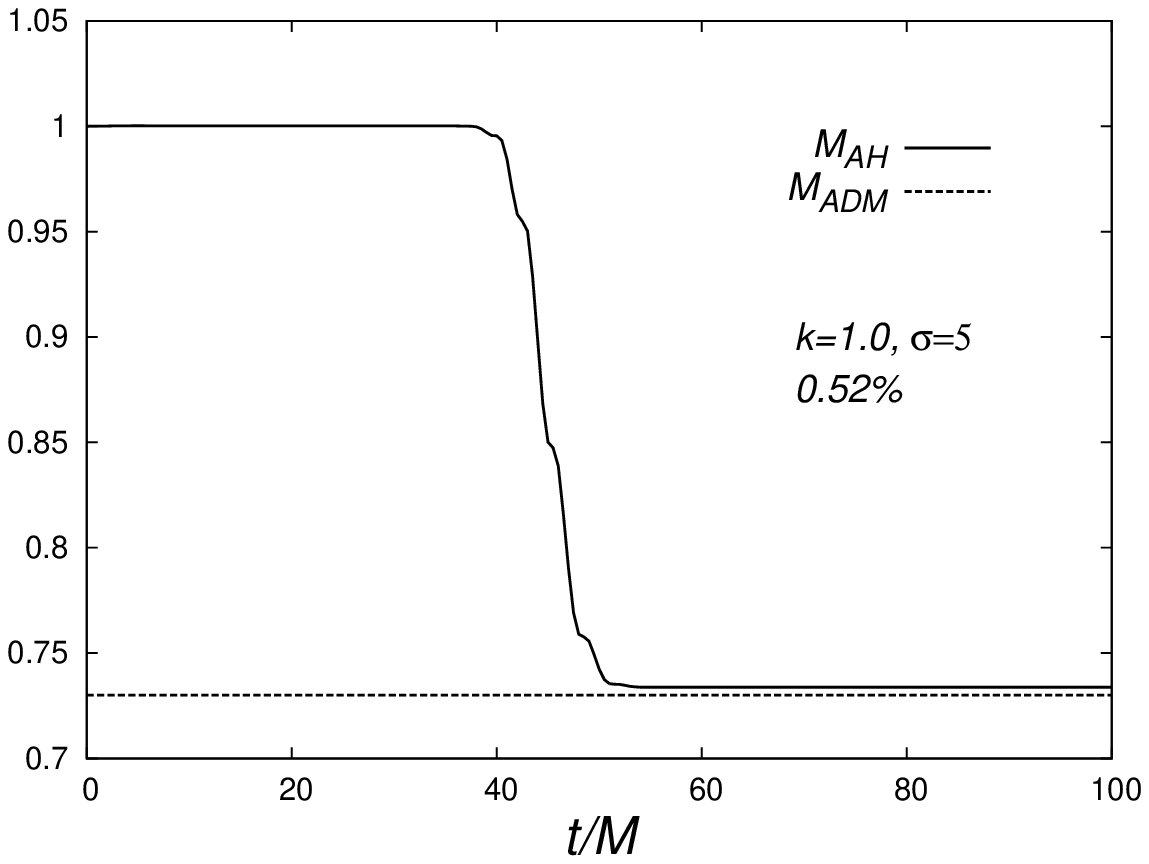}
\includegraphics[width=4.3cm]{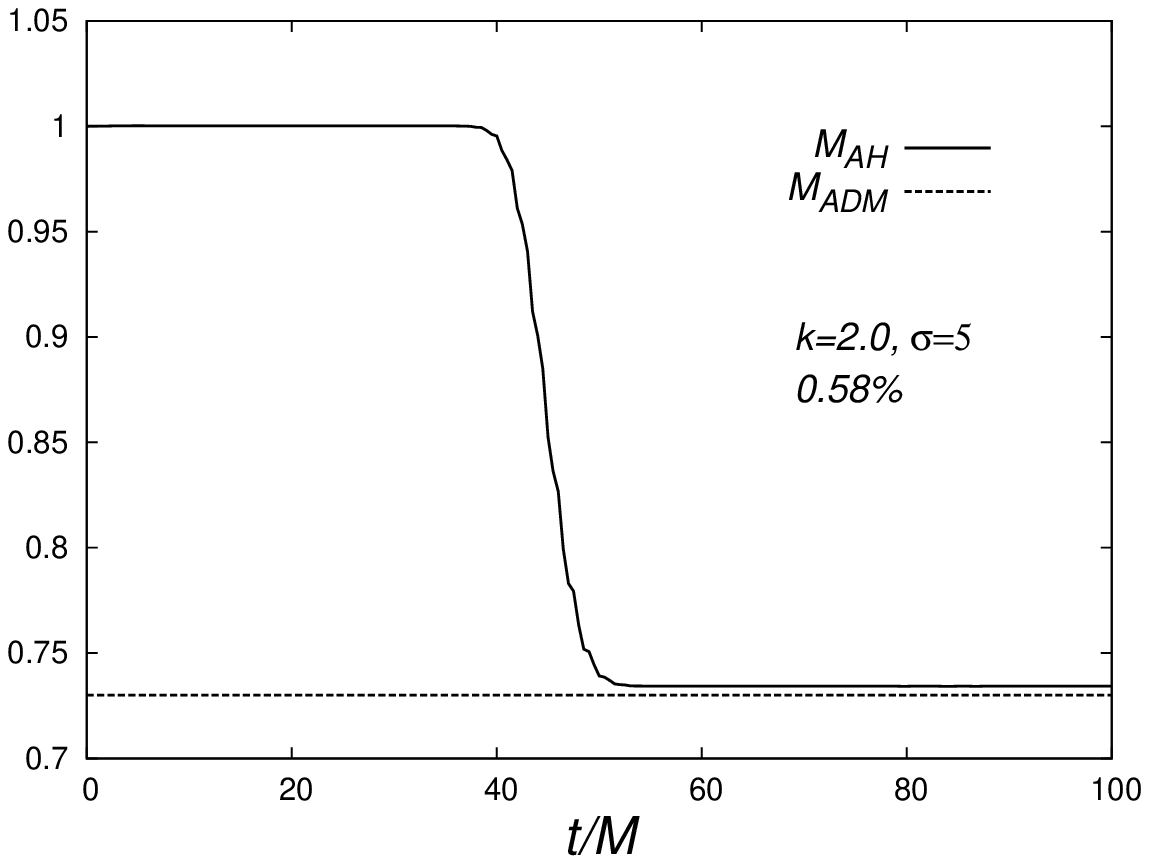}
\includegraphics[width=4.3cm]{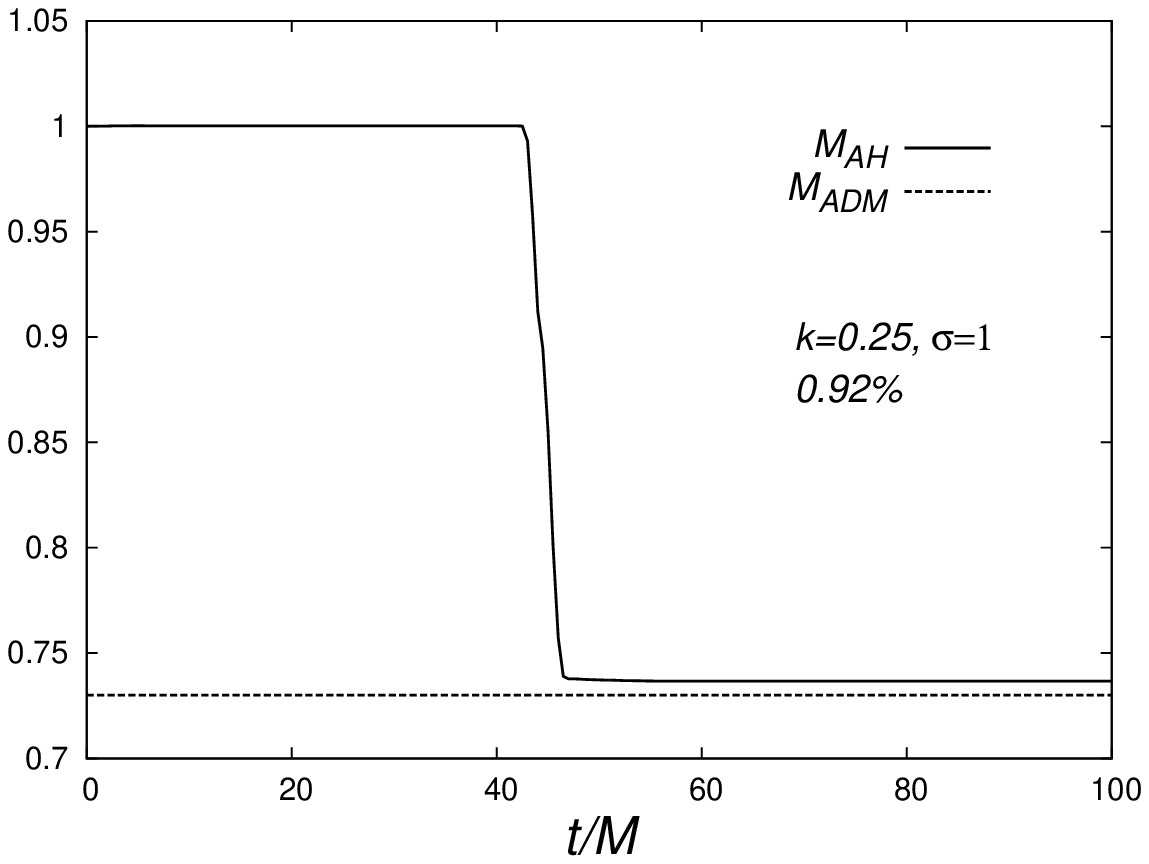}
\includegraphics[width=4.3cm]{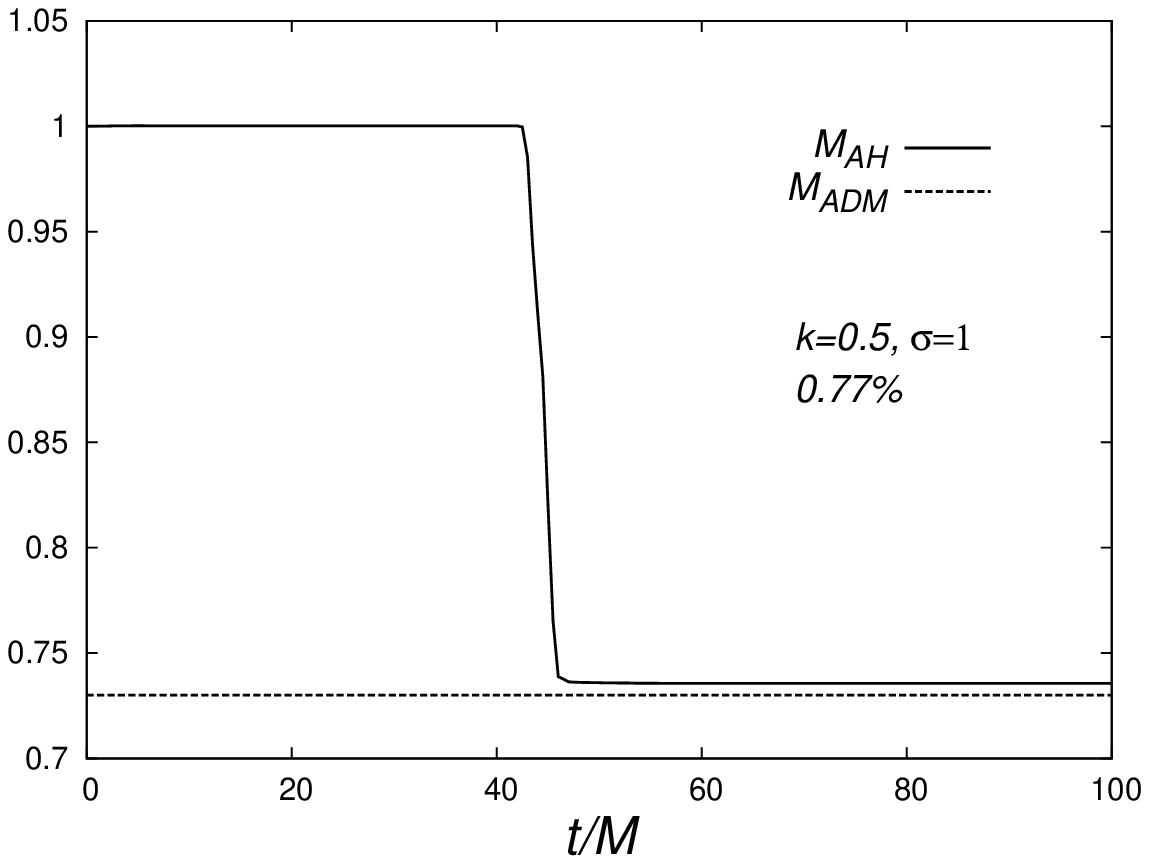}
\includegraphics[width=4.3cm]{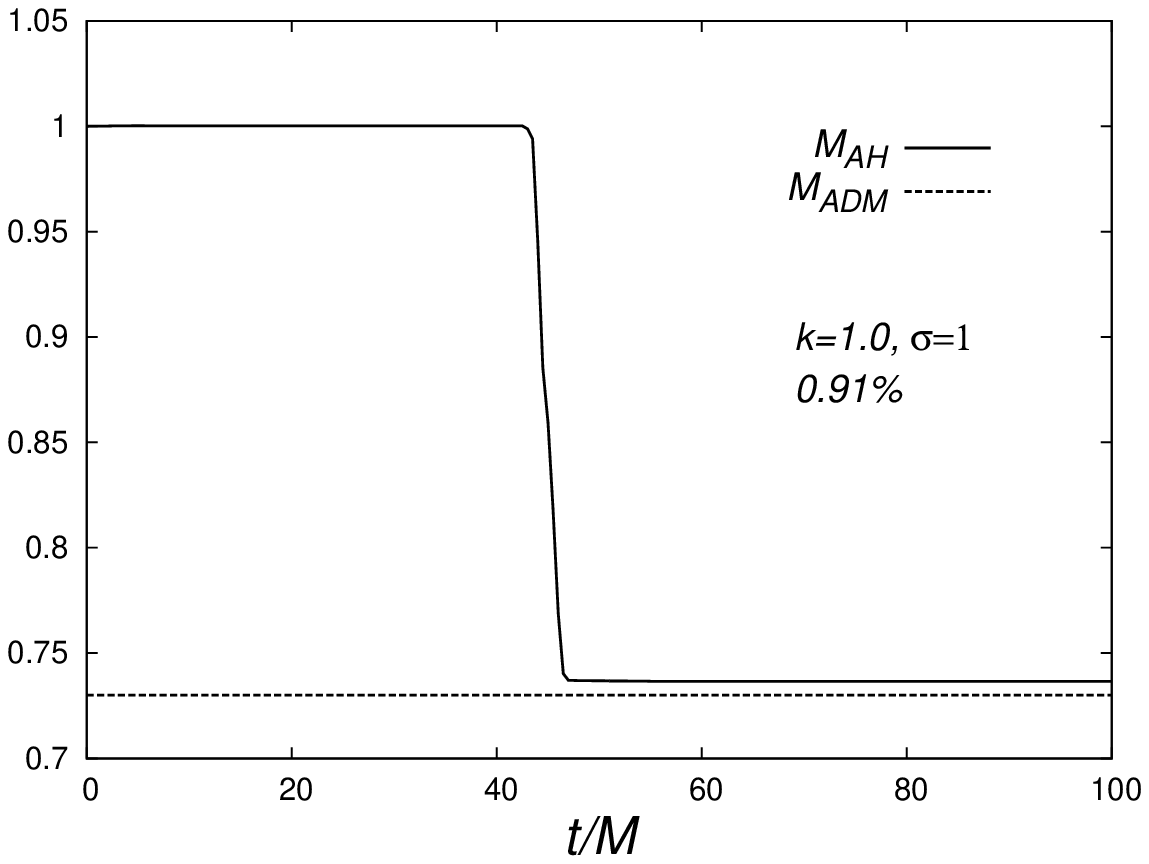}
\includegraphics[width=4.3cm]{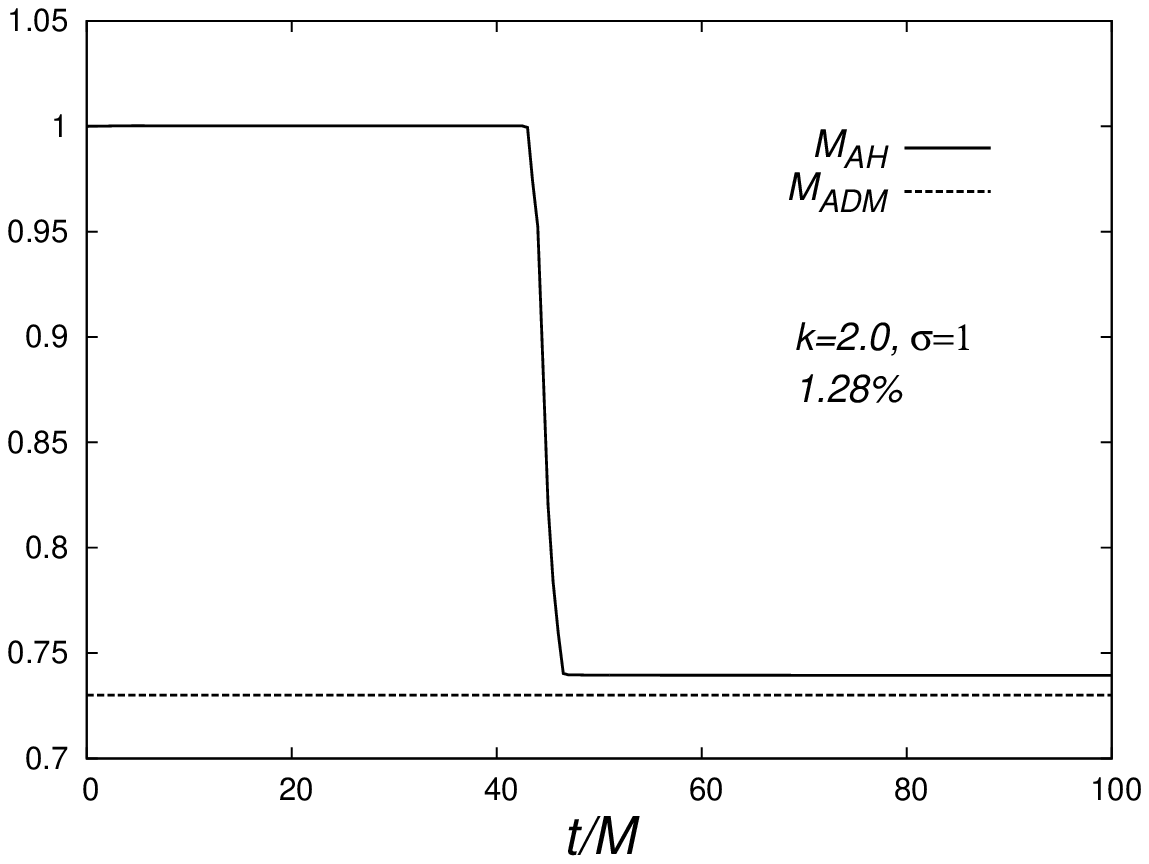}
\caption{\label{fig:paramspace} $M_{AH}$ and $M_{ADM}$ in time for various combinations of the parameters of the initial scalar field $k=0.25,0.5,1.2M$ and the width of the wave packet $\sigma=1,5,10 M$ in units where $M$ is the initial mass of the black hole alone. We also show the percentage of scalar field that is not accreted by the black hole.}
\end{figure*}

Our main results are summarized in Figure \ref{fig:paramspace}, where we show the black hole apparent horizon mass $M_{AH}$ and $M_{ADM}$ during sufficient time for the system to approach stationary mass values.

In the cases where $M_{AH}$ approaches the value of $M_{ADM}$, the scalar field is nearly completely accreted by the black hole. In all the cases we explored there is always a difference between these two masses. The difference in the two values corresponds to the scalar field that was not accreted and remained surrounding the black hole distributed across the spatial domain.


\begin{figure}[htp]
\includegraphics[width=7.5cm]{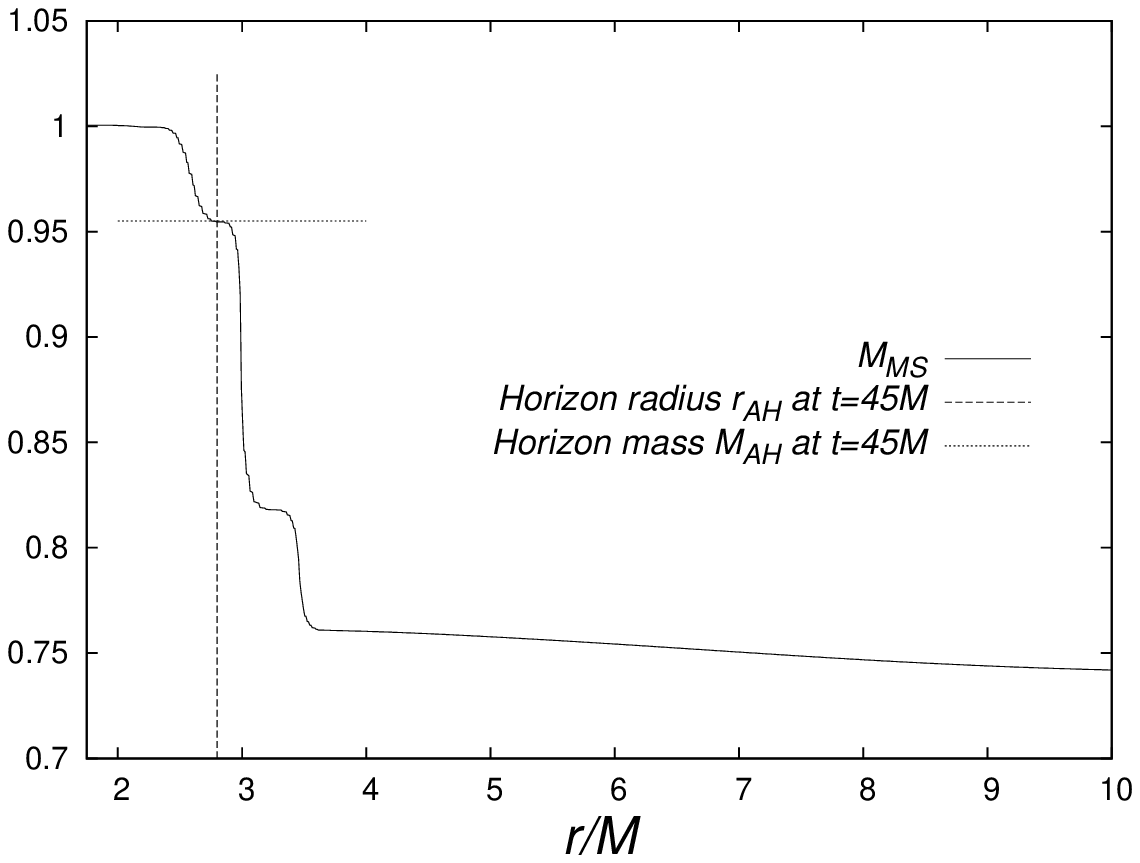}
\includegraphics[width=7.5cm]{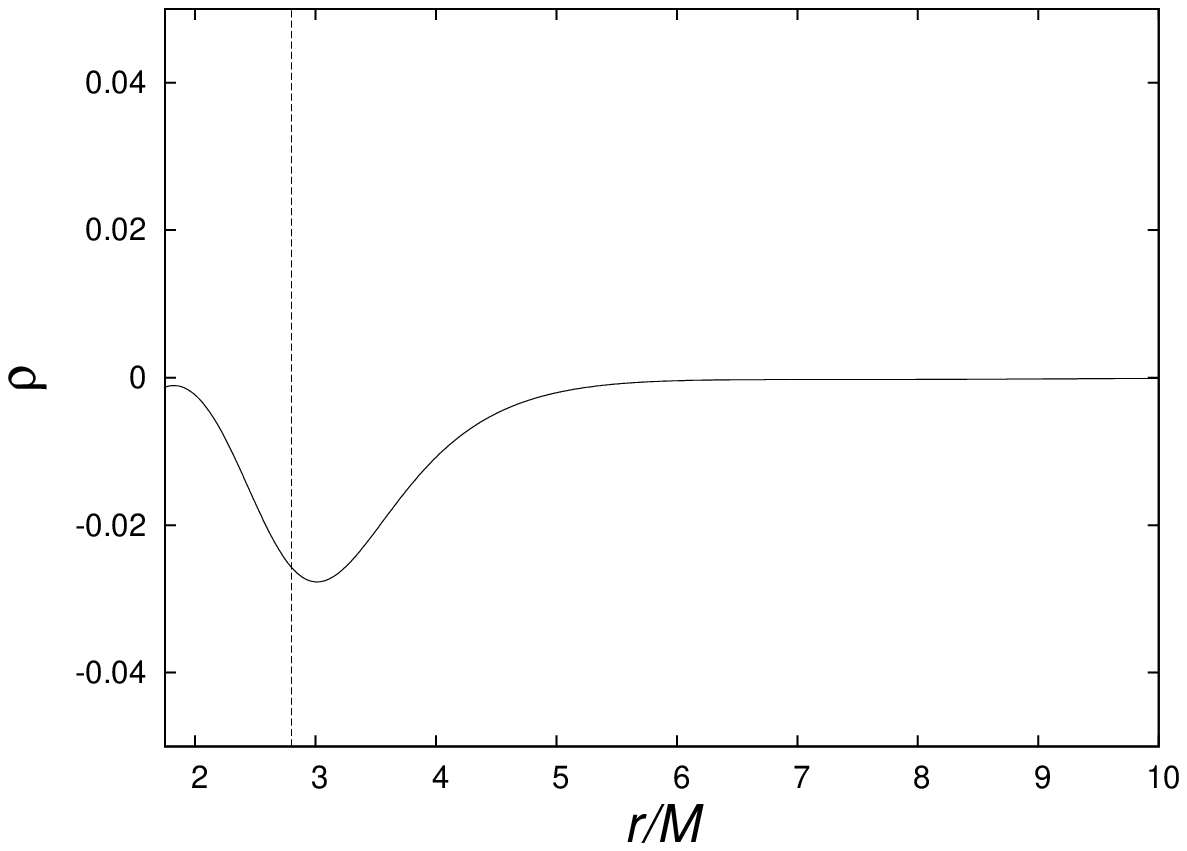}
\caption{\label{fig:snapshot} (Top) We show $M_{MS}$ at $t=45M$ for the parameters $k=0.25$ and $\sigma=1$, where we show that $M_{MS}$ approaches the value of the apparent horizon mass, as expected near the horizon. (Bottom) We show the scalar field density near the horizon, which distributes around the black hole. The vertical line indicates the location of the apparent horizon $r_{AH}$ at $t=45M$ in the $r$-coordinate.}
\end{figure}

In each case we indicate the percentage of scalar field that was not accreted by the black hole. In most cases the non-accreted scalar field ranges around the 1\% of the initial incoming pulse mass. Specifically, for a small sigma $\sigma=M$, the wave packet is concentrated in a thin shell and the accretion is very efficient independently of the wave number $k$. However, for $\sigma=5,10M$, that is, for a width packet bigger than the black hole radius, the accretion is not as efficient anymore for all the values of $k$. It happens that the accretion is inefficient when $k=0.25$, or equivalently when the wavelength of the scalar field is large. We found two representative examples of this case, one with 3.3\% ($\sigma=5M$) and another one with 4.1\% ($\sigma=10M$).

What these results show is that the combination of a thick wave packet together with a large wavelength, avoid the scalar field to be fully accreted.

In order to acquire some insight about the space-time masses relation and as a second check of our calculations, we show in Fig. \ref{fig:snapshot} a snapshot at $t=45M$ of the Misner-Sharp mass together with the black hole mass and radius of the apparent horizon, and compare its value at the apparent horizon surface. We also show in Fig. 3 the distribution of the scalar field density $\rho=T_{\mu\nu}n^{\mu}n^{\nu}$ after the apparent horizon mass has stabilized.

\section{Conclusions}
\label{sec:conclusions}

We have analyzed the spherical accretion of a massless phantom scalar field in order to investigate the possibility that the accretion depends on the properties of the scalar field distribution.

We have found similar results to those found for the accretion of a regular scalar field on a fixed space-time background \cite{Urena2011} and for the full non-linear accretion \cite{GuzmanLora2012}. These results indicate that the wave number $k$ is associated with the amount of scalar field absorbed by the black hole, resulting in that the smaller the $k$ the bigger the amount of scalar field not entering the black hole, as long as $\sigma$ is bigger than the BH size. This is also an example of a possible interpretation of the scalar field wavelength as a sort of length scale for the interaction or non-interaction with a black hole \cite{UrenaLiddle}. Our results can provide some bounds, for instance in the study of vanishing black holes near a Big Rip scenario \cite{Ref86}.

This indicates that the effect of reducing the black hole mass through the accretion of a phantom scalar field is not as general as suggested in \cite{Ref86} using a fixed space-time analysis and in \cite{GG1} where full General Relativity has been used to attempt evaporating black holes.

An interesting question would be that of the distribution of the scalar field remaining outside the black hole. So far the distribution problem has only been explored in the fixed space-time regime for a usual scalar field \cite{ccs2}, an effect that has been dubbed scalar field wigs. What we show in this paper is that, using full General Relativity, a black hole can also wear a halo of phantom scalar field.

Among the interesting scenarios to analyze in the future there is the massive case ($V\ne0$) and the accretion efficiency in terms of the mass and wave number relations. Also it would be interesting to analyze the accretion of scalar fields in other theories, including different flavors of Galileon fields \cite{BabichevReview}.


\section*{Acknowledgments}

This research is partly supported by grants CIC-UMSNH 4.9 and 4.23, CONACyT 258726 (Fondo Sectorial de Investigaci\'on para la Educaci\'on). The simulations were carried out in the IFM Draco cluster funded by  CONACyT 106466.


\end{document}